\documentclass[aps,pre,twocolumn,showpacs,superscriptaddress,reprint]{revtex4}
\usepackage{amsmath}
\usepackage{amsfonts}
\usepackage{amssymb}
\usepackage{epsfig}
\usepackage{graphicx}
\usepackage[usenames,dvipsnames]{color}
\usepackage{ulem}

\newcommand{\sad}{\mbox{\scriptsize sad}}
\newcommand{\therm}{\mbox{\scriptsize therm}}
\newcommand{\thresh}{\mbox{\scriptsize th}}

\newcommand{\ave}[1]{\langle {#1} \rangle}

\begin{document}

\title{Mean-field dynamic criticality and geometric transition in the Gaussian core model}

\author{Daniele Coslovich}
\affiliation{Laboratoire Charles Coulomb, UMR 5221 CNRS-Universit\'e de Montpellier, Montpellier, France}
\author{Atsushi Ikeda}
\email[Corresponding author: ]{atsushi.ikeda@fukui.kyoto-u.ac.jp}
\affiliation{Fukui Institute for Fundamental Chemistry, Kyoto University, Kyoto, Japan}
\thanks{D. Coslovich and A. Ikeda contributed equally to this work.}

\author{Kunimasa Miyazaki}
\affiliation{Department of Physics, Nagoya University, Nagoya, Japan}

\date{\today}

\begin{abstract}
We use molecular dynamics simulations to investigate dynamic heterogeneities and the potential energy landscape of the Gaussian core model (GCM). 
Despite the nearly Gaussian statistics of particles' displacements, the GCM exhibits giant dynamic heterogeneities close to the dynamic transition temperature.
The divergence of the four-point susceptibility is quantitatively well described by the inhomogeneous version of the Mode-Coupling theory.
Furthermore, the potential energy landscape of the GCM is characterized by a geometric transition and large energy barriers, as expected from the lack of activated, hopping dynamics.
These observations demonstrate that all major features of mean-field dynamic criticality can be observed in a physically sound, three-dimensional model.
\end{abstract}


\maketitle

\section{Introduction}

Supercooled liquids are characterized by collective dynamic
fluctuations, known as dynamic heterogeneities, which occur over longer time- and length-scales as the glass transition temperature $T_g$ is approached.
At the molecular scale, these fluctuations imply the correlated motion of an increasingly large number of molecules as relaxation slows down.
To quantify dynamic heterogeneities, a general formalism based on multi-point dynamic correlations was developed over the last years~\cite{dhbook}.
In particular, the four-point dynamic susceptibility $\chi_4(t)$ allows one to evaluate the amplitude of the dynamic fluctuations in numerical simulations~\cite{ensemble} and, at the cost of some approximations, in experiments~\cite{berthier_science}.

Despite these advances, predicting the temperature evolution of dynamic heterogeneities %for a given material 
remains a big challenge and none of the theories proposed so far is conclusive, as they describe experimental and numerical results equally well or poorly~\cite{cavagnareview,glassreview}.
Amongst them, the mode-coupling theory (MCT) is known to be a microscopic and first principles theory of the glass transition~\cite{gotze}. 
MCT was initially formulated as a theory of caging in liquids and focused on two-point correlators.
A recent generalization of MCT to inhomogeneous systems (IMCT) enables one to evaluate multi-point correlation functions and make quantitative predictions for dynamic heterogeneities~\cite{imct}. 
Within this framework, both relaxation times and dynamic fluctuations diverge
algebraically at the dynamic transition temperature $T_c$.
These divergences are however ``avoided'' in real glass-formers:
The dynamics at low temperature is instead governed by thermal activation, with a distinct super-Arrhenius temperature dependence. 
In this regime, dynamic heterogeneities are expected as a manifestation of cooperatively rearranging regions, or mosaics, as predicted by the classic Adam-Gibbs scenario~\cite{adamgibbs}.

The random first order transition theory (RFOT), which was originally inspired by mean-field models of spin glasses, integrates 
these two apparently distinct scenarios~\cite{rfot,biroli2009}. 
According to RFOT, the dynamic transition predicted by MCT corresponds to the trapping of the system in one of the basins of its rugged free energy landscape.
In the mean-field limit, the dynamics is completely frozen-in at $T_c$,
whereas in finite dimensions the transition is rounded by
thermal activation 
and becomes a mere crossover.
Seen from this perspective, the dynamic transition marks a change in the topology of the landscape (also known as ``geometric transition''~\cite{cavagna}): above $T_c$, the system mostly resides close to saddles, whereas below $T_c$ it is trapped close to local minima. 
The corresponding real space picture implies the existence of two
distinct length scales characterizing dynamic heterogeneities: a dynamic one, which grows algebraically 
on approaching $T_c$, 
and a static one corresponding, crudely speaking, to the size of the mosaics.
  
Experimental data at ambient~\cite{stickel,martinez-garcia} and
high pressure~\cite{casalini}, as well as recent simulation
results~\cite{kob_nature} hint at a dynamic crossover at a temperature
higher than $T_g$.
Around the crossover, however, other physical mechanisms, such as dynamic facilitation~\cite{keys} and/or local structure
formation~\cite{tanaka}, may play an important role and compete with the mean-field scenario.
Indeed, several predictions of the IMCT/mean-field framework remain so far undetected. 
First, the fitted power law exponents  describing the growth of $\chi_4$ and the dynamic correlation length do not completely agree with the ones predicted by IMCT~\cite{karmakar}.
Moreover, according to MCT and IMCT, dynamic fluctuations grow near $T_c$ but the single-particle dynamics remains essentially Gaussian.
This counter-intuitive behavior is absent in standard glass-formers, for which $\chi_4$ and the non-Gaussian parameter $\alpha_2$ are typically correlated and grow concomitantly as the dynamics slows down~\cite{vogel}.
Finally, the saddles that become marginally stable at $T_c$ should be
delocalized~\cite{biroli2009}, but no trace of such extended modes was
detected in the potential energy surface of common
glass-formers~\cite{daniele1}.

In this paper, we put the mean-field scenario to a crucial test by studying the approach to the dynamic transition in the Gaussian core model (GCM)~\cite{stillingergcm}. 
Conventional model glass formers, such as Lennard-Jones and hard sphere fluids, are characterized by short-range, harshly repulsive potentials. 
In the GCM, instead, particles interact via an ultra-soft repulsive potential 
$v(r) = \epsilon e^{-(r/\sigma)^2}$~\cite{stillingergcm,likos}, whose \textit{tail} plays a key role at high density. 
At sufficiently high density, the 
GCM becomes a good glass-former and its average dynamics is well described by MCT~\cite{gcm,gcmdynamics}.
Simulations of the high-density GCM are computationally expensive, due to the large number of neighbors each particle is interacting with.
Therefore, characterizing dynamic fluctuations and the energy landscape of this model requires a major overhaul of the numerical protocol compared to previous studies~\cite{gcm,gcmdynamics}.
By employing state-of-art molecular dynamics simulations on graphics processor units (GPU), we demonstrate that the GCM provides a striking incarnation of mean-field dynamic criticality in a three-dimensional system and provides a solid reference to understand how and when the mean-field scenario is washed out in more common glass-formers. 

\section{Methods}\label{sec:methods}

We use molecular dynamics simulations in the NVT ensemble with a
Nose-Hoover thermostat to study $N=4000$ monodisperse GCM
particles. 
The potential is cut and shifted and smoothed at 
$r_c=4.5\sigma$ with the XPLOR cutoff~\cite{hoomdurl}, which ensures continuity of the force at the cutoff.
For the energy landscape analysis we choose a smaller system ($N=2000$).
In the following, we will use $\sigma$, $10^{-6}\epsilon$, and
$10^{-6}\epsilon/k_B$ as units of the length, energy, and temperature, respectively.
We focus on supercooled fluids along the isochore $\rho=2.0$, for which the thermodynamically stable state is the BCC crystal (for $T<8.2$)~\cite{gcm,gcmstatic}.
The crystallization kinetics is very slow and virtually negligible in our simulations~\cite{gcm,gcmdynamics}. 
We note that the density is much higher than $\rho \approx 0.24$, the re-entrant melting density of GCM~\cite{lang,prestipino,gcmstatic}. 
In the low density limit, the GCM approaches asymptotyically the three-dimensional hard sphere model~\cite{stillingergcm}. 
In this regime, the physics of model differs markedly from the one observed at high density.
To ensure good statistics on the four-point dynamic susceptibility,  we performed four independent production runs
for each temperature and the simulation time for each trajectory was
typically 100 times longer than the structural relaxation time $\tau_\alpha$ (see below). 

Computer simulations of the GCM in the high density, supercooled regime are computationally demanding and require particular care.
Due to quasi-long range nature of the potential and the large statistics needed to evaluate $\chi_4$, the simulation protocol must be efficient.
Moreover, in contrast to systems with harshly repulsive interactions, the force summation is not dominated by the first coordination shell and involves a large number of atoms. 
This, in turn, rises obvious issues of numerical accuracy.
To tackle these issues, we employed the HOOMD simulation package~\cite{hoomdurl, anderson, hoomd2}, a state-of-art molecular dynamics code running on graphic processor units (GPU) with double precision arithmetics. 
HOOMD is currently one of the few simulation codes running entirely on GPU that allows double precision evaluation of \textit{both} the forces and the integration step. 
We checked the results obtained with HOOMD against those of in-house simulation codes running on CPU
over the available temperature range and found them to be nicely consistent.
An initial batch of simulations was performed using the RUMD simulation package~\cite{rumd_arxiv}.
RUMD implements the force calculation in single precision, which turns out to be insufficient for the GCM at high density.
In fact, small but systematic differences between the simulations performed with RUMD and with our in-house codes appeared at sufficiently low temperature, in both NVE and NVT ensembles. 
These data sets were not retained in the analysis.

The stationary points of the potential energy surface (PES) were located using standard numerical strategies adopted in earlier studies on LJ mixtures~\cite{cavagna}.
To locate local minima and saddle points we minimized the total potential energy $U$ and the total force squared
\begin{eqnarray}
W=\frac{1}{N}\sum_{i=1}^N f_i^2
\end{eqnarray}
where $f_i$ is the norm of the force vector on particle $i$
respectively, using the LBFGS minimization algorithm~\cite{liu__1989}.
For each studied temperature, we considered 80 independent configurations as starting points of our $U$- and $W$-minimizations.

Due to the large system size and the long range cutoff, $U$- and $W$-minimizations for the GCM are technically more difficult than for LJ mixtures.
To reduce the computational burden without biasing the results, we used a smaller system size ($N=2000$) compared to the one used to characterize the dynamics ($N=4000$).
We note that for $N=2000$ particle at $\rho=2.0$ the box length $L=10$ is only slightly larger than 
twice the cutoff distance.

Due to softness of the potential and the large number of interacting particles, minimizations require high numerical precision as well as a smooth cut-off scheme.
Thanks to the XPLOR cutoff, we could converge $U$-minimizations to values of the mean squared total force of order $W\approx 10^{-13}$.
For most of the configurations we located this way, the dynamical matrix contained no imaginary modes.
$U$-minimizations that did not converge to true minima  were therefore discarded from the analysis.
The fraction of discarded configurations ranged from less than 10\% (close to $T_c$) to about 20\% (at the highest temperatures).
We note however that inclusion of such spurious configurations does not alter the average inherent structure energy within statistical noise.
To try to improve the convergence, we replaced the XPLOR cutoff by a smoother, cubic interpolation scheme~\cite{cavagna} during the minimization. 
We found that the cubic splined cutoff did not improve appreciably the accuracy of the  minimizations compared to the XPLOR cutoff. 
This indicates that the main numerical difficulty lies in the high dimensionality of the system, which makes the minimization problem ill-conditioned.

$W$-minimizations are known to locate \textit{true} saddle points only rarely~\cite{sampoli__2003}.
Most of the points located through such a procedure are, in fact, \textit{quasi}-saddles, i.e. local minima of $W$ with finite $W$ values.
Our minimization algorithm locates configurations with $W$ of the order $10^{-11}$, which is close to but still larger than the threshold we used for local minima ($W=10^{-13}$).
We conclude that the points located by our $W$-minimizations should be considered as quasi-saddles.
Previous studies showed however that the statistical properties of quasi-saddles and true saddles are practically indistinguishable~\cite{sampoli__2003}.
On this basis, we assumed the equivalence of these two types of points in our analysis.

\section{Results}

\begin{figure}
\includegraphics[width=\linewidth]{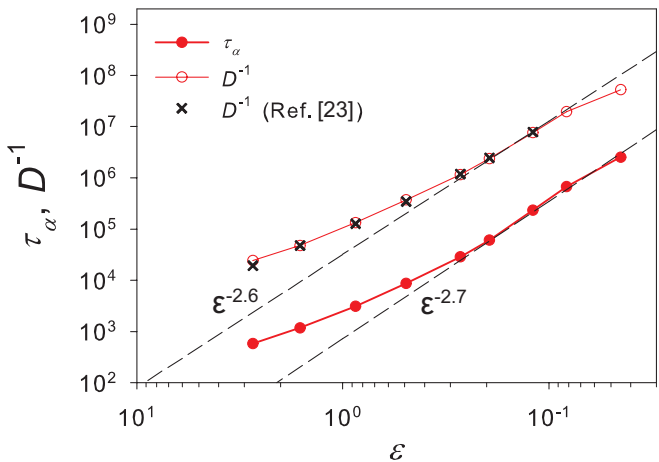}
\caption{\label{si:tau}
Relaxation times $\tau_{\alpha}$ (red filled circles) and inverse diffusion constants $D^{-1}$ (red open circles) 
of the GCM as a function of the reduced temperature $\varepsilon = T/T_c - 1$.  
The $D^{-1}$ of the GCM obtained in Ref.~\cite{gcm} (black crosses) are also plotted. 
The dashed lines are fits to $\tau_{\alpha} \approx \varepsilon^{-\gamma}$ with $\gamma=2.7$ and
$D^{-1} \approx \varepsilon^{-\gamma}$ with $\gamma=2.6$.
}
\end{figure}

Before investigating dynamic heterogeneities, we study the relaxation dynamics of the model close to $T_c$.
We measured the time-dependent overlap function defined by 
$\ave{\hat{F}(t)} = \ave{ N^{-1} \sum_i \Theta (|\Delta \vec{R}_i(t)| - a)}$, 
where $\Delta \vec{R}_i(t)$ is the displacement of the $i$-th particle
in the time interval $t$, $\Theta(x)$ is the Heavyside's step function, and we choose
$a=0.3$, which maximizes the four-point susceptibility defined later. 
$\ave{\hat{F}(t)}$ gives the average fraction of particles which moved more than $a$ after a time $t$.
The relaxation time $\tau_{\alpha}$ is defined by $\ave{\hat{F}(t=\tau_{\alpha})}=e^{-1}$. 
Our equilibrium simulations extend down to $T=2.8$, which corresponds to a relaxation time 4 
times longer than those accessible to previous simulations~\cite{gcm,gcmdynamics}.
The increase of $\tau_\alpha$ becomes non-Arrhenius
around the onset temperature $T_o=5.0$ %(hereafter called the onset temperature $T_o$) 
and displays a power law behavior 
$\tau_\alpha \sim \varepsilon^{-\gamma}$ at low temperature. %with an exponent $\gamma=2.7$.

{Figure~\ref{si:tau} compares the relaxation times $\tau_{\alpha}$ and the diffusion constants $D$ obtained in this work to those of Ref.~\cite{gcm}. 
The data are plotted against the reduced temperature $\varepsilon = T/T_c - 1$. 
The mode-coupling temperature $T_c$ was determined following the standard fitting procedure: we fixed the exponent
$\gamma=2.7$, obtained by solving the MCT equation for the GCM, and then determined $T_c$ by linear regression of
$\tau_{\alpha}^{-1/\gamma}$ against $T$ in the range  $T \leq 3.2$. 
The so obtained value of $T_c = 2.68$ is only 20\% smaller than the theoretical prediction ($T_c=3.2$), 
which should be contrasted with more than 100\% for the Kob-Andersen
Lennard-Jones (KA) mixture~\cite{ka, grzegorzka}.  
We point out that the MCT power-law fit works for a wider range of $\varepsilon$, viz. down to $\varepsilon=0.037$, than in the KA mixture, for which deviations are already apparent around $\varepsilon=0.1$~\cite{ka, grzegorzka}.
Furthermore, the new simulations allow us to detect a slight discrepancy between the exponents that fit the relaxation times ($\gamma=2.7$) and the inverse diffusion constants ($\gamma=2.6$).
In the latter case, we fixed the critical temperature $T_c$ to the one obtained from the relaxation times analysis and only adjusted the power law amplitude and exponent.
These minor deviations from MCT predictions should be contrasted to those found in other models, which are typically around 20--25\%.
We note that the small discrepancy between $\tau$ and $1/D$ observed in the GCM could be explained even within the framework of the MCT itself, as discussed in Ref.~\cite{biroli__2004}.
}

\begin{figure}[t]
\psfig{file=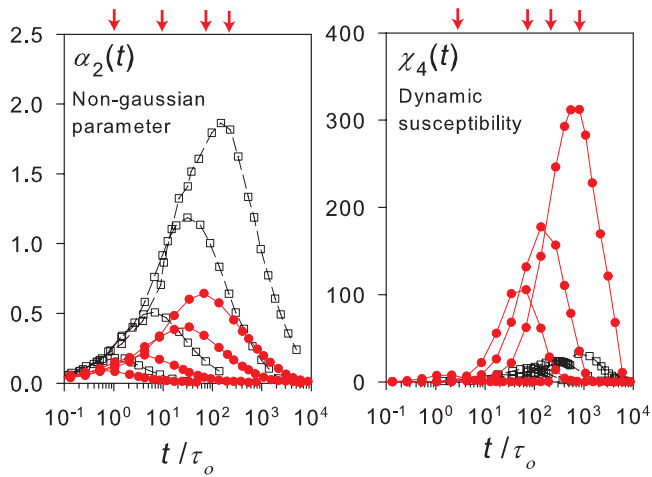,width=8.5cm,clip}
\caption{\label{fig1}
Left: non-Gaussian parameter $\alpha_2(t)$ of the GCM (red circles) at $T=5.0$, 3.4, 3.0 and 2.9, 
and of type A particles of the KA mixture~\cite{ka} (empty squares) at $T=1.0$, 0.6, 0.5 and 0.466. 
Right: four-point susceptibility $\chi_4(t)$ of the GCM (red circles) at $T=4.0$, 3.0, 2.9 and 2.8, 
and of the KA~\cite{karmakar} (empty squares) at $T=0.55$, 0.5, 0.47 and 0.45. 
Downward arrows indicate the corresponding relaxation times $\tau_{\alpha}$ of the GCM.
Time $t$ is scaled by $\tau_{o}$, the relaxation time at the onset temperature $T_o$.  
($T_o = 5.0$ and $\tau_o = 3100$ for the GCM, and $T_o=1.0$ and $\tau_o = 15$ for the KA in the unit of Ref.~\cite{ka}) 
All data sets in the right panel are obtained from molecular dynamics simulations in the NVT ensemble. 
}
\end{figure}

\begin{figure}
 \includegraphics[width=1.0\columnwidth]{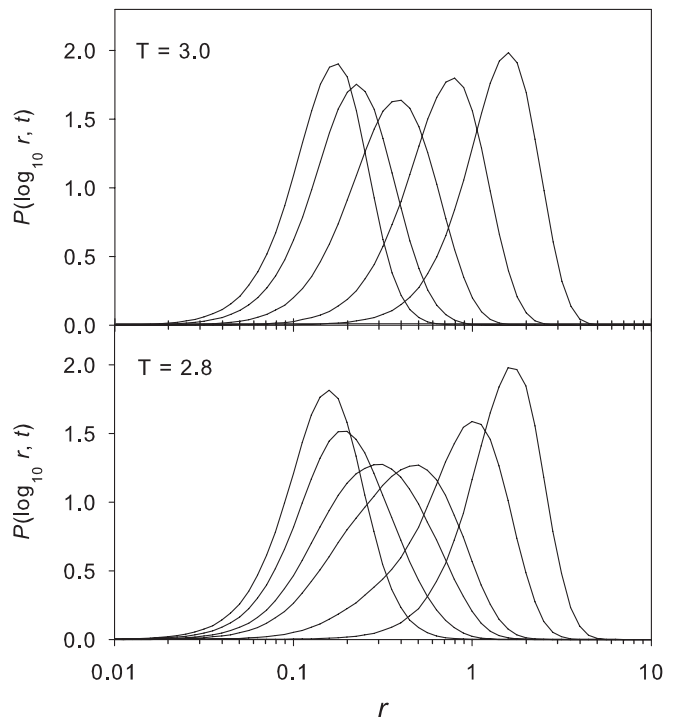} 
\caption{\label{si:dis}
Probability distribution of the logarithm of the particle displacements $P(\log_{10} r, t) \equiv (\ln 10)4 \pi r^3 G_s(r,t)$.
Top: results for $T = 3.0$. From left to right, $t/\tau_o$ = 4, 17, 68, 271 and 1082, where $\tau_{\alpha} / \tau_o = 75$. 
Bottom: results for $T = 2.8$. 
From left to right, $t/\tau_o$ = 34, 135, 406, 812, 3112 and 8659, where $\tau_{\alpha} / \tau_o = 813$. 
}
\end{figure}

To characterize dynamic heterogeneities in the GCM, we consider two different observables.
First, we evaluate the non-Gaussian parameter $\alpha_2(t) = 3\ave{\Delta R(t)^4}/5\ave{\Delta R(t)^2}^2 -1$,  
which quantifies how much the particles' displacements deviates from a Gaussian distribution. 
In the left panel of Fig.~\ref{fig1}, we plot $\alpha_2(t)$ for the KA mixture
and the GCM for similar relaxation time windows. 
$\alpha_2(t)$ of the KA mixture grows rapidly as the temperature is decreased whereas the growth of $\alpha_2(t)$ of the GCM is moderate.
This observation is consistent with the shape of the van Hove functions 
$G_s(r,t) = \ave{ N^{-1} \sum_i \delta (|\Delta R_i(t)| - r)}$ near $t=\tau_{\alpha}$. 
{Figure~\ref{si:dis} shows the probability distribution of the logarithm of single-particle displacements, 
which is related to the van Hove functions as $P(\log_{10} r, t) \equiv (\ln 10)4 \pi r^3 G_s(r,t)$, of the GCM at $T=3.0$ and 2.8. 
For standard model glass formers, the shape of this distribution function is bimodal close to $T_c$, 
which proves the existence of hopping-like motion~\cite{grzegorzka}. 
For the GCM, by contrast, the distribution functions remains unimodal even at the lowest investigated temperature. 
The absence of hopping-like motion in the GCM is consistent with our analysis of the potential energy landscape (see below) which indicates that thermally activated relaxation is strongly suppressed.
We conclude that the  distribution of single-particle mobility in the GCM is homogeneous, even when the slow dynamics is well developed.}
The shape of $G_s(r,t)$ of the KA mixture is bimodal, indicating
coexistence of highly mobile and immobile particles~\cite{grzegorzka},  while that of the GCM remains unimodal %and Gaussian-shaped 
even at the lowest temperature. %(see Fig.~2 in Supplemental Information). 

Next, we evaluate the four-point dynamic susceptibility defined by 
$\chi_4(t) = N[\ave{\hat{F}(t)^2} - \ave{\hat{F}(t)}^2]$, which
quantifies the cooperative motion of particles in fluids~\cite{glotzer,karmakar}.
Strikingly, the trend is now reversed as shown in the right panel of
Fig.~\ref{fig1}: 
$\chi_4(t)$ grows far more strongly in the GCM than in the KA mixture.
Around $T_c$, the maximum of $\chi_4(t)$, called herein $\chi_4^*$, in the GCM is 
one order of magnitude larger than that of the KA mixture.
Note that while the dynamic susceptibility depends in general on the statistical ensemble, 
this result holds also in the ensemble where all global variables are allowed to fluctuate (see Appendix).
Thus dynamic fluctuations in the GCM are significantly larger than those in other standard models around $T_c$.

The opposite trends of $\alpha_2(t)$  and $\chi_4(t)$ may look
contradicting at a first glance. %, since they are usually positively correlated~\cite{vogel}.
This 
implies that the nature of dynamic heterogeneities of the GCM is qualitatively different from other conventional glass formers.
Large values of $\alpha_2(t)$ %in the KA mixture is 
are a direct consequence of the large displacements of individual mobile particles and do not 
necessarily reflect the extent of cooperative motion. 
On the other hand, $\chi_4(t)$ is defined as the variance of the
overlap function, which does not depend on how far the mobile particles
have moved, and it is thus more sensitive to the net cooperativity.
Therefore, the suppression of $\alpha_2(t)$ and the concomitant enhancement of $\chi_4(t)$ in the GCM
imply {\it slight but spatially extended modulations of the mobility field}.

\begin{figure*}[!ht]
\includegraphics[width=.8\linewidth]{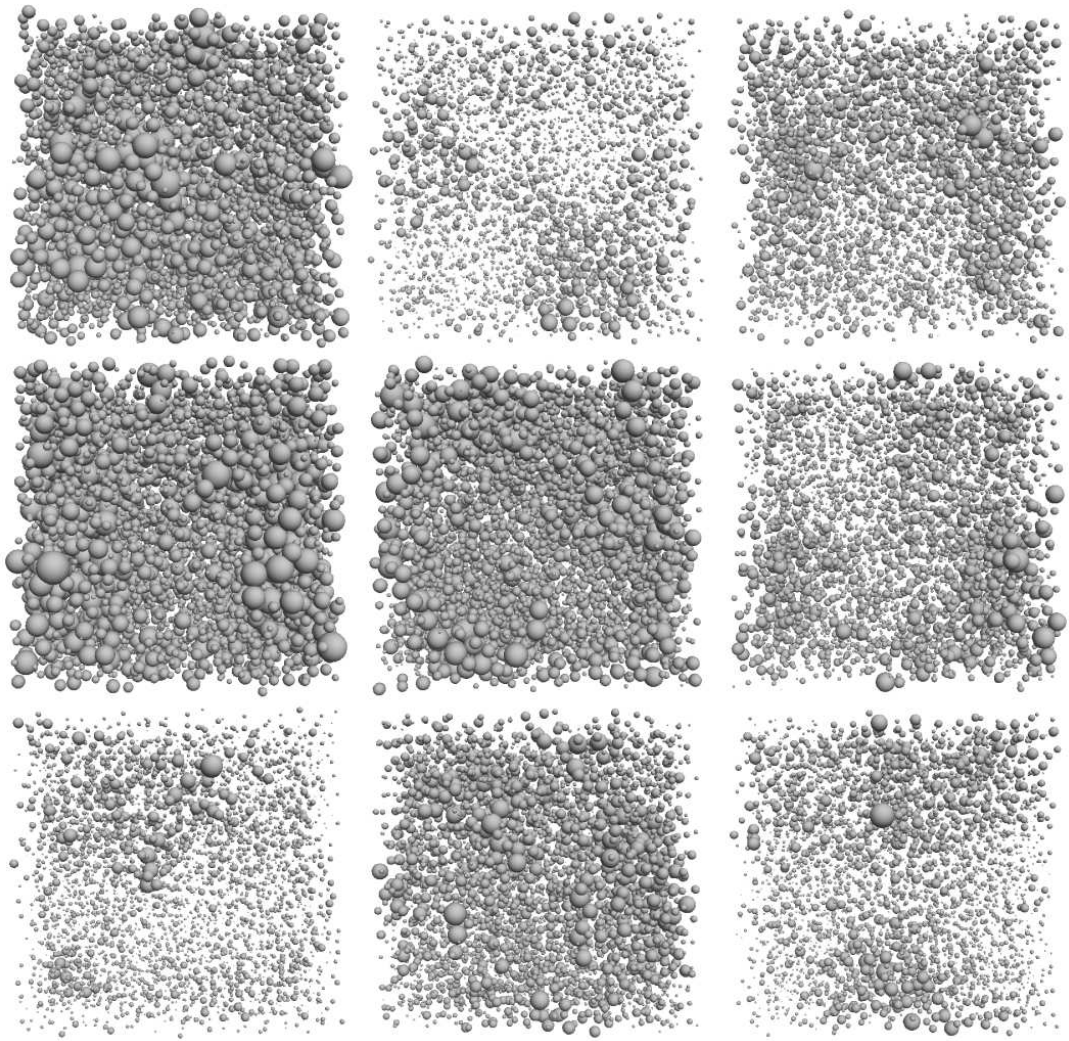}
\caption{\label{si:snap}
Typical snapshots of the particles' mobility field at a temperature $T=2.9$. Particles are shown as spheres of radius proportional to the mobility $\delta r(t,t_0)$ after a time $t=4\times{10}^{5}\approx \tau_{\alpha}$. The time origins $t_0$ at which the configurations are shown are separated by at least 10 structural relaxation times.
}
\end{figure*}

These inferences are confirmed by visual inspection of the mobility field close to the dynamical critical temperature.
To visulalize the giant dynamic fluctuations that give rise to the increase of $\chi_4$ close to the dynamic transition, we evaluate the mobility of the particles after time $t$ as $\delta r_i(t,t_0) = |\vec{r}_i(t+t_0) - \vec{r}_i(t_0)|$.
In Figure~\ref{si:snap} we show typical snapshots of the mobility field at $T=2.9$, close to the dynamic transition.
At this temperature, the maximum of the dynamic susceptibility has reached about 200.
The radii of the spheres are proportional to the particles' displacements after a time $t=4\times10^{5}\approx \tau_{\alpha}$.
We clearly see that the mobility field is characterized by extended regions of either mobile or immobile particles, with very smooth variations over space.
Note that no coarse-graining or time averaging is involved in our calculation of $\delta r$.
Visual inspection suggests that the correlation length is comparable to the system size at this temperature.
This is in turn compatible with the existence of finite size effects around and below this temperature.

\begin{figure}[t]
\psfig{file=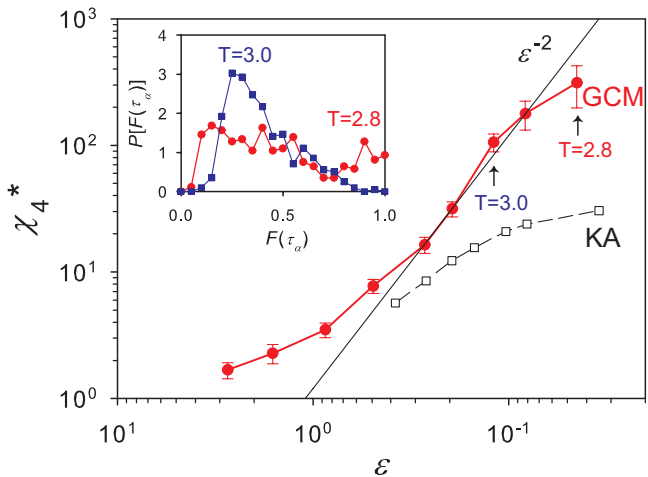,width=8.5cm,clip}
\caption{\label{fig2}
Maximum value of the dynamic susceptibility, $\chi_{4}^{\ast}$,  
against the reduced temperature $\varepsilon =T/T_c - 1$ in the GCM (red circles) and
in the KA mixture~\cite{karmakar} (empty squares), where
$T_c = 2.68$ for the GCM and $T_c = 0.435$ for the KA.
The solid line is the IMCT prediction, $\chi_4^{\ast} \sim \varepsilon^{-2}$. 
The inset shows the histogram of the overlap between two configurations $\hat{F}(t)$
with the time interval $t=\tau_{\alpha}$ for the GCM at $T=3.0$ and 2.8. 
}
\end{figure}

The coexistence of giant dynamic heterogeneities and Gaussian-like
dynamics is a strong evidence of mean-field dynamic criticality.
According to MCT and IMCT, 
the amplitude of $\alpha_2(t)$ remain small~\cite{fuchs1998}, whereas 
$\chi_4^{\ast}$ diverges on approaching $T_c$ with an exponent
that depends on both microscopic dynamics and statistical ensemble~\cite{imct,ensemble,grzegorzdh}.  
For molecular dynamics simulations in the NVT ensemble, the
susceptibility is predicted to diverge as $\chi_4^{\ast} \sim \varepsilon^{-2}$. 
Figure~\ref{fig2} shows the dependence of $\chi_4^{\ast}$ on $\varepsilon$. 
The data for the KA mixture are not well described by IMCT, whereas
for the GCM $\chi_{4}^{\ast}$ follows the power law predicted by the theory over the range of temperature $2.8<T\leq 3.4$. 
We emphasize that in Fig.~\ref{fig2} the critical temperatures $T_c$ are those obtained by fitting the relaxation times data.
Our data demonstrate that $\tau$, $1/D$, and $\chi_4^*$ can be fitted by power laws over
comparable $\varepsilon$ ranges and using the appropriate set of MCT exponents.

Deviations from the IMCT prediction are observed only at the
 lowest temperature in Fig.~\ref{fig2}. 
This deviation is most likely due to finite size effects, which will naturally appear if $\chi_4(t)$ has a genuine divergence, rather than to the crossover between MCT and activated regimes.
In the inset of Fig.~\ref{fig2}, we plot the histogram of the
overlap $\hat{F}(t)$ for the GCM at $t=\tau_{\alpha}$.
By definition, the mean value of the histogram is $\ave{\hat{F}(t=\tau_{\alpha})}= e^{-1}$.
Indeed, the histogram at $T=3.0$ is unimodal with the mean value of
$e^{-1}$, whereas it becomes very broad at $T=2.8$, suggesting 
that a correlation length becomes comparable to the system size.  
We note that such strong finite size effects were not observed in the KA mixture~\cite{karmakar,berthier_fs}. 

We now proceed to discuss the dynamics of the GCM from the perspective of the energy landscape~\cite{goldstein,sastry}.
According to the RFOT scenario, the MCT crossover should be accompanied by a
``geometric'' transition in the energy landscape~\cite{cavagnareview}:
in mean-field, the unstable modes separating the free energy minima become marginally
stable as $T_c$ is approached and eventually disappear below a certain
threshold energy~\cite{cavagna}. 
In finite dimensions, remnants of this geometrical transition
should be visible in the potential energy landscape  around the
MCT crossover: Above $T_c$ the system relaxes mostly via unstable soft modes,
while activated relaxation over energy barriers takes over below
$T_c$. 
The crossover between the two regimes takes place at an energy threshold
$e_{\thresh}$, below which the system resides mostly close to local minima of
the potential energy instead of stationary points of arbitrary order. 
Early numerical simulations of the KA mixture gave support to
this scenario and found a crossover temperature $T_{\thresh}$ very close to
$T_c$~\cite{cavagna,angelani}, but were later on called into question~\cite{berthier_garrahan,wales_doye}.

\begin{figure}[t]
\psfig{file=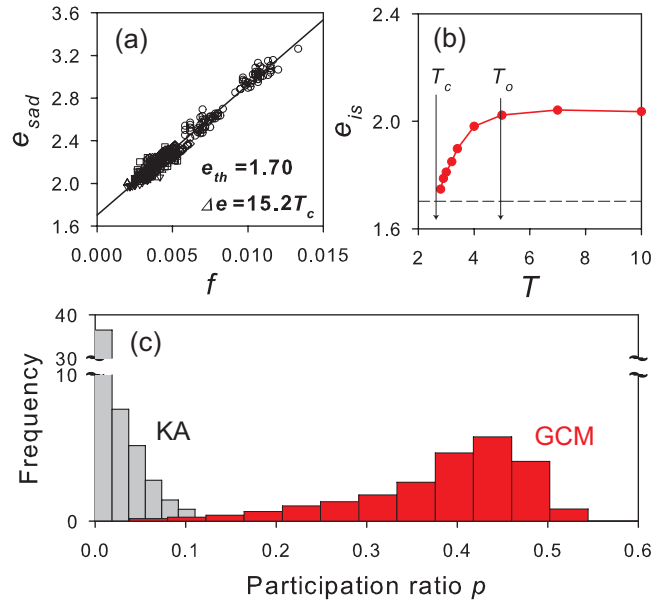,width=8.5cm,clip}
\caption{\label{fig3}
Analysis of the potential energy landscape of the GCM. 
(a) Energy of saddle points $e_{\sad}$ as a function of the fraction of unstable modes $f$. 
The solid line is a linear fit $e_{\sad} = e_{\thresh} + (3\Delta e) f$ with $e_{\thresh} = 1.70$ and $\Delta e/T_c = 15.2$. 
(b) The inherent structure energy $e_{is}$ vs. temperature. 
The horizontal dashed line indicates $e_{\thresh} = 1.70$. 
Vertical arrows indicate the MCT temperature $T_c = 2.68$ and the onset temperature $T_o = 5.0$.
(c) Distributions of the participation ratio of unstable modes of the KA mixture ($N=2000$)
 at $T =0.45$ and of the GCM at $T = 2.9$. 
}
\end{figure}

To determine the statistics of stationary points, we located minima and saddles by applying the LBFGS
minimization algorithm to the total potential energy $U$ and to the total force $W$
respectively~\cite{daniele1} (see Sec.~\ref{sec:methods}). 
Although local minima of $W$ do not necessarily correspond to true stationary points~\cite{wales_doye}, $W$-minimizations yield a fairly robust measurement of the energy threshold $e_{\thresh}$ and of the typical energy barriers~\cite{angelani}.
Figure~\ref{fig3}(a) shows the saddle point energy,
$e_{\sad}$, as obtained from $W$-minimizations, against
the fraction of unstable modes $f$ found in the spectrum of the
dynamical matrix 
\footnote{The potential energy is displayed after subtracting the
trivial ``uniform'' contribution 
$\epsilon(\rho \sigma^3 \pi^{3/2}-1)/2$ out. See~\cite{gcmstatic}.}. 
As in other models, these two quantifies are roughly linearly related.
From a linear fit $e_{\sad} = e_{\thresh} + (3\Delta e) f$ we extract the threshold energy $e_{\thresh} = 1.70$.
Comparing this with the temperature dependence of the energy of minima $e_{m}$ (Fig.~\ref{fig3} (b)), 
we obtain a threshold temperature $T_{\thresh} \approx 2.7$ in excellent agreement with $T_c$.
The slope $\Delta e$ gives an estimate of the average barrier height,
since it is the energy cost to increase by one the order of a stationary
point~\cite{cavagna}. 
We found $\Delta e/T_c = 15.2$ in the GCM, which is appreciably larger $\Delta e/T_c \approx 10$ observed in other model glass formers~\cite{cavagnareview,coslovich_network}.
We estimate that the increased barrier height hampers the activated relaxation by a factor $\exp(15)/\exp(10) \approx 150$. 
As the activated relaxation channels are strongly suppressed, the
MCT-like critical behavior dominates the slow dynamics of the model~\cite{cavagna}.

Further support for the geometric transition scenario is provided by the analysis of the mode localization.
It has been argued that the unstable directions that disappear at the dynamic transition are 
\textit{delocalized} in the mean-field scenario~\cite{biroli2009}. 
This contrasts with the typical observation that unstable modes of
common glass-formers become increasingly localized as $T$ decreases~\cite{daniele2}. 
To evaluate the spatial localization of the unstable modes, we calculate their participation ratio $p$ on a per-mode basis 
\begin{eqnarray}
p(\omega) = \left\langle \left( N \sum_i (\vec{e}_i(\omega) \cdot \vec{e}_i(\omega))^2 \right)^{-1} \right\rangle
\end{eqnarray}
where $\vec{e}_i(\omega)$ is the displacement of particle $i$ along the eigenvector corresponding to the eigenfrequency $\omega$.
In Fig.~\ref{fig3}(c) we plot the distribution of the
participation ratio of unstable modes %of the KA mixture and the GCM sampled 
around $T_c$.
As in various conventional glass formers~\cite{daniele2}, the unstable
modes of the KA mixture have small participation ratios and are therefore spatially localized. 
By contrast, the distribution is broader in the GCM and considerably
shifted towards larger value of $p$ ($0.6$ for plane waves, 1 for
completely delocalized modes). 
Visual inspection suggests that these modes have a complex spatial structure, in which some groups of particles undergo cooperative motions, while others display incoherent displacements.
While determining the precise nature of the modes in the GCM would

While determining the precise nature of the modes in the GCM would require an analysis of $N$-dependence of the spectrum, 
and thus of the mobility edge~\cite{bembenek}, our analysis suggests that the giant 
dynamic heterogeneities of the GCM might indeed be associated to these extended unstable modes.
In the KA mixture, instead, dynamic heterogeneities build up through dynamic facilitation of localized elementary rearrangements~\cite{keys}, 
which might be related to modes localized outside locally stable domains~\cite{daniele2}. 

\section{Conclusions}

In summary, we presented evidence that the mean-field scenario predicting diverging dynamic fluctuations and a geometric transition close to the dynamic transition can be observed in a three-dimensional model system at sufficiently high density. 
Our results shed new light on the physical nature of dynamic heterogeneities predicted by 
IMCT. 
The approach to the dynamic transition is accompanied by rapidly growing
dynamic correlations and by the disappearance of extended unstable modes
of the potential energy landscape. 
We identify a clear fingerprint of mean-field dynamic criticality,
that is, the coexistence of the strong dynamic fluctuations and nearly
Gaussian distribution of a single particle displacement.
Which finite-dimensional features (e.g. locally preferred structures,
dynamic facilitation) mask the
mean-field physics in actual supercooled liquids is a question that need
to be addressed in future studies. 

\acknowledgments
We thank two anonymous reviewers for useful remarks and constructive criticisms on a previous version of the manuscript. 
We acknowledge the HPCLR 
Center of Competence in High-Performance Computing of Languedoc-Roussillon (France)
and ACCMS of Kyoto University (Japan) for allocation of CPU time. 
A.I. acknowledges the financial support from JSPS postdoctoral fellowship for research abroad and KAKENHI No. 26887021. 
K.M. is supported by KAKENHI No.
24340098,
25103005,  
and 25000002. 

\appendix
\section{Checks on ensemble dependence}

\begin{figure}
\includegraphics[width=\linewidth]{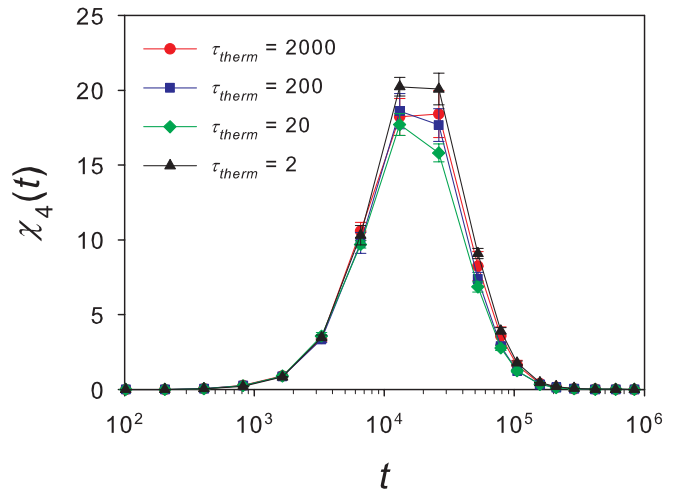}
\caption{\label{si:thermo}
Effect of the thermostat relaxation time $\tau_{\therm}$ on the dynamic susceptibility $\chi_4$ at $T=3.4$ in the GCM. 
The values of the thermostat relaxation time are indicated in the figure caption. Most of our simulations were carried out using $\tau_{\therm} = 200$. 
}
\end{figure}

Most of our simulations were performed in the NVT ensemble by means of the Nose-Hoover thermostat. 
It is important therefore to make sure that the choice of the thermostat relaxation time
did not affect the dynamic quantities, in particular the dynamic susceptibility.
In Fig.~\ref{si:thermo} we show $\chi_4(t)$ evaluated using four different thermostat relaxation times $\tau_{\therm}=2$, 20, 200, 2000. 
In our simulations we choose $\tau_{\therm}=200$.
The differences between $\chi_4(t)$ obtained with different $\tau_{\therm}$ remain within statistical uncertainties, thus our choice of $\tau_{\therm}$ does not affect the results. 

\begin{figure}
\includegraphics[width=\linewidth]{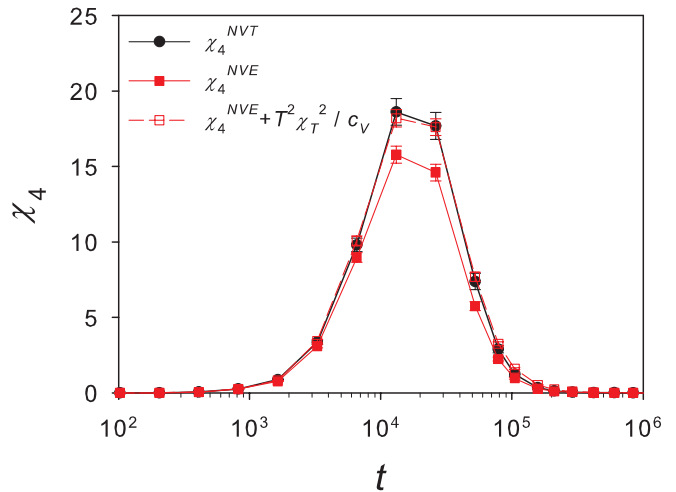}
\caption{\label{si:nve}
Statistical ensemble dependence of the dynamic susceptibility in the GCM. 
The dynamic susceptibility measured in the NVT ensemble ($\chi_4^{NVT}$, circles) and in the NVE ensemble ($\chi_4^{NVE}$, filled squares) are plotted for $T=3.4$.
Open squares indicate $\chi_4^{NVE} + T^2 \chi_T^2 /c_V$, which must be identical to $\chi_4^{NVT}$~\cite{ensemble}.  
}
\end{figure}

As is well established, the dynamic susceptibility depends in general on the statistical ensemble~\cite{ensemble}.
Since the thermostat relaxation time $\tau_{\therm}$ is long enough such that $\chi_4(t)$ is independent from $\tau_{\therm}$ itself (Fig.~\ref{si:thermo}),  
the evaluated $\chi_4(t)$ can be safely considered as the dynamic susceptibility in the NVT ensemble~\cite{Flenner_Szamel_2013}.
It is possible to transform the dynamic susceptibility between ensembles by means of well-known transformation formulas.
In particular, the dynamic susceptibility in the NVT ensemble is related to the one in the NVE ensemble by $\chi_4^{NVT} = \chi_4^{NVE} + T^2 \chi_T^2 /c_V$, where $\chi_T = \partial \ave{\hat{F}(t)}/\partial T$ and $c_V$ is the specific heat at constant volume. 
This, in turn, provides us with an internal check of our calculations. 
In Fig.~\ref{si:nve} we compare $\chi_4^{NVE}$, $\chi_4^{NVT}$, and $\chi_4^{NVE} + T^2 \chi_T^2 /c_V$. 
For this calculation, we averaged over 16 independent simulation runs for each ensemble to improve the statistical accuracy. 
We see that $\chi_4^{NVE} + T^2 \chi_T^2 /c_V$ agrees with $\chi_4^{NVT}$ within error bars.
We also note that the difference between $\chi_4^{NVE}$ and $\chi_4^{NVT}$ is small at this temperature. 

Finally, we point out that $\chi_4^{NVT}$ obtained from our NVT simulation is essentially the same as 
the total dynamic susceptibility $\chi_4^\textrm{total}$, defined as the volume integral of the four point correlator. 
$\chi_4^\textrm{total}$ can be expressed as the sum of $\chi_4^{NVT}$ and $\chi_4|_{\delta N}$, the fluctuation of dynamics originating from particle number fluctuations.

\begin{figure}
\includegraphics[width=\linewidth]{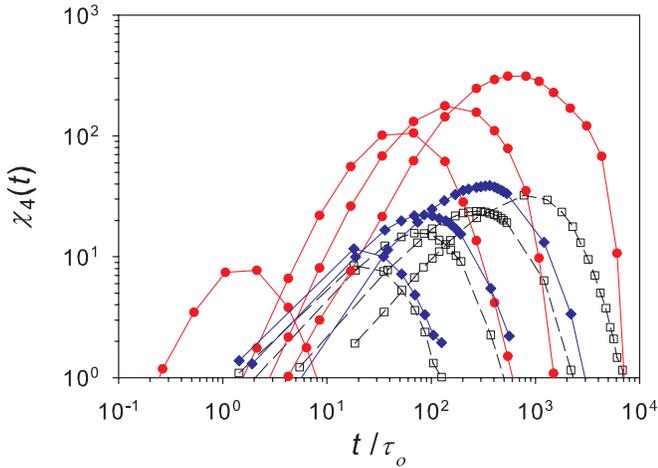}
\caption{\label{si:chi4}
Log-log plot of $\chi_4(t)$ in the NVT ensemble for the GCM (red circles) 
at $T=4.0$, 3.0, 2.9 and 2.8 (from left to right) 
and for the KA mixture (empty squares) at $T=0.55$, 0.5, 0.47 and 0.45 (from left to right). 
Blue diamonds indicate the full dynamic susceptibility $\chi_4^\textrm{total}$ of
the KA model at $T=0.55$, 0.5 and 0.47 (from left to right).
}
\end{figure}

To assess the role of $\chi_4|_{\delta N}$ quantitatively, we calculated it explicitly for both GCM and KA models. 
For a binary mixture, $\chi_4|_{\delta N}$ can be expressed as~\cite{Flenner_Szamel_2013} 
\begin{eqnarray}
\chi_4 |_{\delta N} &=& \chi_{\rho}^2 H_1 + \chi_{\rho} \chi_{c} H_2 + \chi_{c}^2 H_3+ \ave{\hat{F}(t)}^2 H_4 \nonumber \\
&&  + \ave{\hat{F}(t)}\chi_{\rho} H_5 + \ave{\hat{F}(t)}\chi_{c} H_6
\label{eq:chiN}
\end{eqnarray}
with 
\begin{eqnarray}
H_1 &=& \rho^2 c S_{11} + 2 \rho^2 \sqrt{c(1-c)} S_{12} + \rho^2 (1-c) S_{22}  \nonumber \\
H_2 &=& 2\rho c(1-c) S_{11} + 2 \rho (1-2c) \sqrt{c(1-c)} S_{12} \nonumber \\
&& - 2\rho c (1-c) S_{22} \nonumber \\
H_3 &=& c(1-c)^2 S_{11} - 2c (1-c) \sqrt{c(1-c)} S_{12} \nonumber \\
&& + c^2 (1-c) S_{22}  \nonumber \\
H_4 &=& c S_{11} + 2 \sqrt{c(1-c)} S_{12} + (1-c) S_{22}  \nonumber \\
H_5 &=& 2\rho c S_{11} + 4\rho  \sqrt{c(1-c)} S_{12} + 2\rho (1-c) S_{22} \nonumber \\
H_6 &=& 2c(1-c) S_{11} + 2 (1-2c) \sqrt{c(1-c)} S_{12} \nonumber \\
&& - 2c (1-c) S_{22}  \nonumber  
\end{eqnarray}
where $\chi_\rho = ( \partial \ave{\hat{F}(t)}/\partial \rho)_c$, $\chi_c = ( \partial \ave{\hat{F}(t)}/\partial c)_{\rho}$, 
$\rho$ is the dimensionless number density and $c$ is the concentration of the species 1 in the binary mixture. 
$S_{nm}$ is the $k \to 0$ limit of the partial structure factor $S_{nm}(k)$. 
The present expressions differ from those in Ref.~\cite{Flenner_Szamel_2013}, 
because we treat $\rho$ (instead of the packing fraction) and $c$ as independent variables. 
We checked that both formulations give the same results for $\chi_4 |_{\delta N}$ in the KA model at $T=0.55$. 
Finally, if one sets $c=1$ in the above expressions, one gets the expression for $\chi_4|_{\delta N}$ for monodisperse systems. 

First we calculated $\chi_4 |_{\delta N}$ for the GCM at $T=3.2$ and
3.4 using the expressions above 
and found it to be of order $10^{-3}$ at these temperatures and thus negligibly small.
This is largely due to the extremely small compressibility (to which $\chi_4 |_{\delta N}$
is proportional, as shown in Eq.~\eqref{eq:chiN}).
Indeed, we found $S(k \to 0) \approx 10^{-6}$ at the temperatures we studied~\cite{gcmstatic}. 
Furthermore, $\chi_\rho^2$  in Eq.~\eqref{eq:chiN} does not diverge faster than $\chi_T^2$. 
This can be rationalized by the fact that $T_c$ scales like the melting temperature as a
function of density, {\sl i.e.} $T_c \sim \exp(-b\rho^{2/3})$ and thus $\chi_\rho$ should grow 
with the same exponent as $\chi_T$.
From these facts, we conclude that $\chi_4^\textrm{total} \approx \chi_{NVT}$ for the GCM.

Next, let us assess $\chi_4|_{\delta N}$ for the KA model.
We performed additional molecular dynamics simulations of the KA model at $T=$ 0.55, 0.5 and 0.47 to
evaluate $\chi_4|_{\delta N}$. 
The system size of these simulations was $N=1000$.
We also performed a simulation with a larger system size ($N=8000$) at $T=0.55$ 
to confirm our extrapolation of $S_{nm}(k)$ in the $k \to 0$ limit. 
The maximum values of  $\chi_4|_{\delta N}$ are 2.5, 7.3, and 15.9 at $T=$ 0.55, 0.5 and 0.47, respectively. 
In Fig.~\ref{si:chi4} we plot the full dynamic susceptibility $\chi_4^\textrm{total} = \chi_4^{NVT}
+ \chi_4 |_{\delta N}$ for the KA model for these three temperatures
and compare them to the results in Fig.~\ref{fig1} of the main text. We point out again that
$\chi_4^\textrm{total}$ should be indistinguishable from $\chi_4^{NVT}$ for the GCM.

From these observations, we conclude that the contrast between the two models around $T_c$ remains sharp even by including the $\chi_4 |_{\delta N}$ term and confirms that dynamic fluctuations are much more pronounced in the GCM than in the KA model.


\begin{thebibliography}{99}
\expandafter\ifx\csname natexlab\endcsname\relax\def\natexlab#1{#1}\fi
\expandafter\ifx\csname bibnamefont\endcsname\relax
  \def\bibnamefont#1{#1}\fi
\expandafter\ifx\csname bibfnamefont\endcsname\relax
  \def\bibfnamefont#1{#1}\fi
\expandafter\ifx\csname citenamefont\endcsname\relax
  \def\citenamefont#1{#1}\fi
\expandafter\ifx\csname url\endcsname\relax
  \def\url#1{\texttt{#1}}\fi
\expandafter\ifx\csname urlprefix\endcsname\relax\def\urlprefix{URL }\fi
\providecommand{\bibinfo}[2]{#2}
\providecommand{\eprint}[2][]{\url{#2}}

\bibitem{dhbook}
{\it Dynamical Heterogeneities in Glasses, Colloids, and Granular Materials}, 
edited by L. Berthier, G. Biroli, J.-P. Bouchaud, L. Cipelletti, and W. van Saarloos 
(Oxford University, New York, 2011).

\bibitem{ensemble}
L. Berthier, G. Biroli, J.-P. Bouchaud, W. Kob, K. Miyazaki and D. R. Reichman, 
J. Chem. Phys. {\bf 126}, 184503 (2007); 
{\it ibid.} {\bf 126}, 184504 (2007).

\bibitem{berthier_science}
L. Berthier, G. Biroli, J.-P. Bouchaud, L. Cipelletti, D. El Masri, D. L'Hote, F. Ladieu, M. Pierno,
Science {\bf 310}, 1797-1800 (2005)

\bibitem{cavagnareview}
A. Cavagna, 
Phys. Rep. {\bf 476}, 51 (2009).

\bibitem{glassreview}
L. Berthier and G. Biroli, 
Rev. Mod. Phys. {\bf 83}, 587-645 (2011).

\bibitem{gotze}
W. G\"otze, 
{\it Complex Dynamics of Glass-Forming Liquids}, 
(Oxford University, Oxford, 2009).

\bibitem{imct}
G. Biroli, J.-P. Bouchaud, K. Miyazaki and D. R. Reichman, 
Phys. Rev. Lett. {\bf 97}, 195701 (2006).

\bibitem{adamgibbs}
G. Adam and J. H. Gibbs, 
J. Chem. Phys. {\bf 43}, 139 (1965).

\bibitem{rfot}
T. R. Kirkpatrick, D. Thirumalai and P. G. Wolynes, 
Phys. Rev. A {\bf 40}, 1045 (1989).

\bibitem{biroli2009}
G. Biroli and J. P. Bouchaud, 
in 
{\it Structural Glasses and Supercooled Liquids}, 
edited by P. G. Wolynes and V. Lubchenko (Wiley, 2012).

\bibitem{cavagna}
T.S. Grigera, A. Cavagna, I. Giardina, and G. Parisi, Phys. Rev. Lett. {\bf 88}, 055502 (2002).

\bibitem{stickel}
F. Stickel, E.W. Fischer, and R. Richert, 
J. Chem. Phys. {\bf 102}, 6251-6257 (1995).

\bibitem{martinez-garcia}
J.C. Martinez-Garcia, J. Martinez-Garcia, S.J. Rzoska, and J. Hulliger,
J. Chem. Phys. {\bf 137}, 064501 (2012).

\bibitem{casalini}
R. Casalini and C.M. Roland, 
Phys. Rev. Lett. {\bf 92}, 245702 (2004).

\bibitem{kob_nature}
W. Kob, S. Roldan-Vargas, and L. Berthier, 
Nature Phys. {\bf 8}, 164 (2012).

\bibitem{keys}
A. S. Keys,  L. O. Hedges, J. P. Garrahan, S. C. Glotzer, and D. Chandler, 
Phys. Rev. X.  {\bf 1}, {021013}, (2011).

\bibitem{tanaka}
H. Tanaka, Eur. Phys. J. E {\bf 35}, 113 (2012)

\bibitem{karmakar}
S. Karmakar, C. Dasgupta and S. Sastry, 
Proc. Natl. Acad. Sci. U.S.A. {\bf 106}, 3675 (2009).

\bibitem{vogel}
M. Vogel and S. C. Glotzer,
Phys. Rev. E {\bf 70}, 061504 (2004).

\bibitem{daniele1}
D. Coslovich and G. Pastore, 
EPL {\bf 75}, 784 (2006).

\bibitem{stillingergcm}
F. H. Stillinger, 
J. Chem. Phys. {\bf 65}, 3968 (1976); 
Phys. Rev. B {\bf 20}, 299 (1979).

\bibitem{likos}
C. N. Likos, 
Soft Matter {\bf 2}, 478 (2006). 

\bibitem{gcm}
A. Ikeda and K. Miyazaki, 
Phys. Rev. Lett. {\bf 106}, 015701 (2011).

\bibitem{gcmdynamics}
A. Ikeda and K. Miyazaki, 
J. Chem. Phys. {\bf 135}, 054901 (2011).

\bibitem{hoomdurl}
http://codeblue.umich.edu/hoomd-blue.

\bibitem{gcmstatic}
A. Ikeda and K. Miyazaki, 
J. Chem. Phys. {\bf 135}, 024901 (2011).

\bibitem{lang}
A. Lang, C. N. Likos, M. Watzlawek and H Lowen, 
J. Phys. Condens. Matter {\bf 12}, 5087 (2000). 

\bibitem{prestipino}
S. Prestipino, F. Saija and P. V. Giaquinta, 
Phys. Rev. E {\bf 71}, 050102 (2005). 

\bibitem{anderson}
J. A. Anderson, C. D. Lorenz and A. Travesset, 
J. Comput. Phys. {\bf 227}, 5342 (2008).

\bibitem{hoomd2}
J. Glaser, T. D. Nguyen, J. A. Anderson, P. Liu, F. Spiga, J. A. Millan, D. C. Morse and S. C. Glotzer. 
Comput. Phys. Commun. {\bf 192} 97 (2015). 


\bibitem[{\citenamefont{Bailey et~al.}(2015)\citenamefont{Bailey, Ingebrigtsen,
  Hansen, Veldhorst, Bøhling, Lemarchand, Olsen, Bacher, Larsen, Dyre
  et~al.}}]{rumd_arxiv}
\bibinfo{author}{\bibfnamefont{N.~P.} \bibnamefont{Bailey}},
  \bibinfo{author}{\bibfnamefont{T.~S.} \bibnamefont{Ingebrigtsen}},
  \bibinfo{author}{\bibfnamefont{J.~S.} \bibnamefont{Hansen}},
  \bibinfo{author}{\bibfnamefont{A.~A.} \bibnamefont{Veldhorst}},
  \bibinfo{author}{\bibfnamefont{L.}~\bibnamefont{B{\o}hling}},
  \bibinfo{author}{\bibfnamefont{C.~A.} \bibnamefont{Lemarchand}},
  \bibinfo{author}{\bibfnamefont{A.~E.} \bibnamefont{Olsen}},
  \bibinfo{author}{\bibfnamefont{A.~K.} \bibnamefont{Bacher}},
  \bibinfo{author}{\bibfnamefont{H.}~\bibnamefont{Larsen}},
  \bibinfo{author}{\bibfnamefont{J.~C.} \bibnamefont{Dyre}},
  \bibnamefont{et~al.}, \bibinfo{journal}{arXiv:1506.05094}
  (\bibinfo{year}{2015}), \urlprefix\url{http://rumd.org}.

\bibitem[{\citenamefont{Liu and Nocedal}(1989)}]{liu__1989}
\bibinfo{author}{\bibfnamefont{D.~C.} \bibnamefont{Liu}} \bibnamefont{and}
  \bibinfo{author}{\bibfnamefont{J.}~\bibnamefont{Nocedal}},
  \bibinfo{journal}{Math. Program.} \textbf{\bibinfo{volume}{45}},
  \bibinfo{pages}{503} (\bibinfo{year}{1989}).

\bibitem[{\citenamefont{Sampoli et~al.}(2003)\citenamefont{Sampoli, Benassi,
  Eramo, Angelani, and Ruocco}}]{sampoli__2003}
\bibinfo{author}{\bibfnamefont{M.}~\bibnamefont{Sampoli}},
  \bibinfo{author}{\bibfnamefont{P.}~\bibnamefont{Benassi}},
  \bibinfo{author}{\bibfnamefont{R.}~\bibnamefont{Eramo}},
  \bibinfo{author}{\bibfnamefont{L.}~\bibnamefont{Angelani}}, \bibnamefont{and}
  \bibinfo{author}{\bibfnamefont{G.}~\bibnamefont{Ruocco}},
  \bibinfo{journal}{J. Phys.: Condens. Matter} \textbf{\bibinfo{volume}{15}},
  \bibinfo{pages}{S1227} (\bibinfo{year}{2003}).
 

\bibitem{ka}
W. Kob and H. C. Andersen, 
Phys. Rev. E {\bf 51}, 4626 (1995); 
{\it ibid.}  {\bf 52}, 4134 (1995).

\bibitem{grzegorzka}
E. Flenner and G. Szamel, 
Phys. Rev. E {\bf 72}, 011205 (2005); 
{\it ibid.} {\bf 72}, 031508 (2005). 

\bibitem[{\citenamefont{Biroli and Bouchaud}(2004)}]{biroli__2004}
\bibinfo{author}{\bibfnamefont{G.}~\bibnamefont{Biroli}} \bibnamefont{and}
  \bibinfo{author}{\bibfnamefont{J.-P.} \bibnamefont{Bouchaud}},
  \bibinfo{journal}{EPL (Europhysics Letters)} \textbf{\bibinfo{volume}{67}},
  \bibinfo{pages}{21} (\bibinfo{year}{2004}).



\bibitem{glotzer}
N. Lacevic, F. W. Starr, T. B. Schroder, and S. C. Glotzer, 
J. Chem. Phys. {\bf 119}, 7372 (2003).

\bibitem{fuchs1998}
M. Fuchs, W. Gotze, and M. R. Mayr, 
Phys. Rev. E  {\bf 58},  3384 (1998).

\bibitem{grzegorzdh}
E. Flenner and G. Szamel, 
Phys. Rev. Lett. {\bf 105}, 217801 (2010).

\bibitem{berthier_fs}
L. Berthier, G. Biroli, D. Coslovich, W. Kob and C. Toninelli, 
Phys. Rev. E {\bf 86}, 031502 (2012).

\bibitem{goldstein}
M. Goldstein, 
J. Chem. Phys. {\bf 51}, 3728 (1969).

\bibitem{sastry}
S. Sastry, P. G. Debenedetti and F. H. Stillinger, 
Nature {\bf 393}, 554 (1998).

\bibitem{angelani}
L. Angelani, R. Di Leonardo, G. Ruocco, A. Scala and F. Sciortino, 
Phys. Rev. Lett. {\bf 85}, 5356 (2000); 
L. Angelani, C. De Michele, G. Ruocco and F. Sciortino, 
J. Chem. Phys. {\bf 121}, 7533 (2004).

\bibitem{wales_doye}
J. P. K. Doye and D. J. Wales
J. Chem. Phys. {\bf 116}, 3777 (2002). 

\bibitem{berthier_garrahan}
L.  Berthier and J. P. Garrahan
Phys. Rev. E {\bf 68}, 041201 (2003). 

\bibitem{coslovich_network}
D. Coslovich and G. Pastore, 
J. Phys. Condens. Matter {\bf 21}, 285107 (2009). 

\bibitem{daniele2}
D. Coslovich and G. Pastore, 
J. Chem. Phys. {\bf 127}, 124504 (2007); 
{\it ibid.} {\bf 127}, 124505 (2007).

\bibitem{bembenek}
S. D. Bembenek and B. B. Laird,
J. Chem. Phys. {\bf 104}, 
5199 (1996).

\bibitem[{\citenamefont{Flenner and Szamel}(2013)}]{Flenner_Szamel_2013}
\bibinfo{author}{\bibfnamefont{E.}~\bibnamefont{Flenner}} \bibnamefont{and}
  \bibinfo{author}{\bibfnamefont{G.}~\bibnamefont{Szamel}},
  \bibinfo{journal}{J. Chem. Phys.} \textbf{\bibinfo{volume}{138}},
  \bibinfo{pages}{12A523} (\bibinfo{year}{2013}).
\end{thebibliography}
\end{document}